\documentclass[authoryear, 10pt, a4paper]{elsarticle}
\usepackage[utf8]{inputenc}
\usepackage[T1]{fontenc}
\usepackage[francais]{babel}

\usepackage[top=2.52cm, bottom=2.52cm, left=2.52cm, right=2.52cm]{geometry}
\usepackage{amsmath,amssymb,amsthm}
\usepackage{graphicx}
\usepackage{natbib}
\usepackage{lineno}
\usepackage{array}
\usepackage{color}
\usepackage{setspace}
\usepackage{multicol}

\begin{document}


\title{A constraint-based framework to study rationality,  competition and cooperation in fisheries} 

\author[rvt1]{Christian Mullon\corref{cor1}}
\ead{Christian.Mullon@ird.fr}
\author[rvt2]{Charles Mullon}

\cortext[cor1]{Corresponding author} 


\address[rvt1]{Unit\'{e} de Recherche MARBEC (UMR212), IRD. France.}
\address[rvt2]{Department of Ecology and Evolution, University of Lausanne, Lausanne, Switzerland}

\begin{abstract} 
In this paper, we present a simplified framework to represent competition, coordination and bargaining in fisheries when they operate under financial and technological constraints. Competition within constraints leads to a particular type of mathematical game in which the strategy choice by one player changes strategy set of the other.
By studying the equilibria and bargaining space of this game when players maximize either profit or fishing capacity, we highlight that differences in financial constraints among players leads to a tougher play, with a reduced bargaining space as the least constrained player can readily exclude another from the competition. The exacerbating effects of constraints on competition are even stronger when players maximize capacity.
We discuss the significance of our results for global ocean governance in a current context characterized by financialization and technological development. We suggest that in order to maximize the chances of fruitful negociations and aim towards a fair sharing of sea resources, it would be helpful to focus on leveling current differences in the constraints faced between competing fishing systems by supporting local financial systems and technological control, before implementing sophisticated economic tools. 
\end{abstract}
 
 
\maketitle

\textbf{Suggested running title:} Rationality,  competition and cooperation in fisheries

\textbf{Keywords} 

\section{Introduction}

Game theory and fisheries science have had a long  intertwined history. The work of \citet{gordon1954economic} on fisheries, which anticipated \citet{Hardin1968Tragedy}'s analysis of the tragedy of the common, was an application of \citet{cournot1838recherches}'s duopoly theory, which itself anticipated a Nash analysis of non-cooperative games (see \citealp{leonard1994reading, myerson1999nash} for historical sketches). Since then, game theory has been widely used for the analysis of fishing systems and as theoretical guidance to organise the exploitation of our seas \citep{sumaila1999review, kaitala2007game, bailey2010application}.

The close relationship between game theory and fisheries is illustrated by the use of the competition for a shared fish resource as an example of the prisoner's dilemma \citep{feeny1996questioning}. When fishing systems maximize their profit, competition leads to a situation in which fishing systems earn less than when they cooperate.  Competition in this case ends in over-exploitation and over-capacity, hence the need to transition towards cooperation and coordination if the long term livelihood of fisheries is to be maintained.

The distinction between competition, cooperation and coordination has been a key point of game theory since John Nash's work \citep{nash1950bargaining, nash1950equilibrium, nash1953two}, which gives a theoretical framework to the basic human practice of bargaining. What has been referred to as Nash's program \citep{binmore1986nash} still structures game theoretical considerations of fisheries systems. The bulk of the work devoted to the use of game theory for fisheries has focused on non-cooperative situations \citep{arnason2000norwegian, lindroos2000nash} or, in the case of cooperation, on the formation of coalitions \citep {lindroos2000nash, pintassilgo2003coalition, lindroos2005coalition}.

Fewer works have articulated fisheries problems in terms of cooperation, coordination and bargaining \citep{armstrong1998sharing}. However, the view that these three processes underlie the relationships among fishing systems at different levels has been endorsed by several authors, who argue that a poor understanding of these process may have hindered our ability to tackle fishing crises \citep{fearon1998bargaining,alcock2002bargaining,munro2009game}. For instance, \cite{alcock2002bargaining} uses game theoretic concepts to argue that the collapse of cod in Canada was at least partly due to a distributional conflict between small-scale and industrial fisheries. The absence of theoretical tools to understand this type of distributional conflict contributed to its delayed resolution and transition towards cooperation, resulting in catch diminutions coming about too late. 

\cite{alcock2002bargaining} points at two culprits for the emergence of  conflict that are evocative of current problems faced by many fisheries and that will guide our research. First, distributional conflict in Canada was caused by small-scale and industrial fisheries suffering different constraints, due in particular to technical and financial differences. Technological progress \citep{squires2013technical} and financialization (i.e. modernization and global expansion of financial tools, \citealp{epstein2005financialization}) have both been playing an increasingly important role in fisheries who increasingly rely on finance to purchase ever more advanced fleets. The two processes feedback to constrain fisheries' scope of action as fisheries become increasingly dependent on technology to increase catch in order to repay finance. 

How technology and financialization influence the competitive and cooperative relationships between fishing systems through the constraints they generate has not been much investigated. Constraint-based game theory is an active field research that has been mostly concerned with games on networks (e.g., \citealp{contreras2004numerical,li2010designing}, see \citealp{krawczyk2005coupled} for an application to environmental problems). Yet, the need for an assessment of the constraining effects of technology and finance is bolstered by many current problems in fisheries other than distributional conflicts, such as the dangers of a rent-maximisation principle in developing countries \citep{bene2010not}, or the economic and ecological stakes associated with deep-sea fishing \citep{norse2012sustainability}. Here, we will show that framing economical problems related to the exploitation of renewable resources within constraints offers a change of perspective on these problems as the effect of an agent's action on another agent translates into diminishing its scope of action, rather than only reducing its income.

The second cause of distributional conflict according to \cite{alcock2002bargaining} are differences in objectives between small-scale and industrial fisheries (such as subsistence vs return on capital). Most of the applications of game theory to fisheries have been based on the idea of profit maximization, and are therefore relevant to one kind of rationality. The variety of modes of rationality at work in social-ecological systems such as fishing systems \citep{berkes2008navigating} can be uncovered by addressing the simple question: who (e.g., boat skipper, firm owner, fishery, fleet, fishery manager, policy maker) maximizes what (e.g., catch, profit, capacity, well-being, pleasure, peace)? Ignoring the variety of modes of rationality in fishing systems has been argued to be potentially damaging for the articulation of efficient fisheries policies \citep{bromley2009abdicating}. In this paper, we will therefore be interested in the effects of rationality on the relationship between fishing systems and in particular, we will study the effects of capacity maximisation in addition to income maximisation.

In view of the above considerations, we propose in this paper a constraint-based framework to study competition, cooperation and rationality  in fisheries systems. The paper is organized as follows. (1) We define a game theoretical framework in order to consider technological and financial constraints, and  games between players with different rationalities. (2) We identify a crucial criterion: the \textit{viability threshold} of a player, which is the maximum yield that a player faced with technological and financial constraints can reach. (3)  We compare the outcomes of games within constraints among competing fishing systems according to whether players maximize income or capital. In particular, we analyze the games' equilibria and their bargaining space and find that the conditions that make it possible to move from cooperation to coordination, to organize bargaining and to reach a fair distribution can be expressed in terms of the viability thresholds of competing players.
 (4) We discuss the relevance of our results to current problems related to ocean governance.

\section{Model}

\subsection{A single fishing system in a constraint-based framework}

 In order to introduce our constraint framework, we first represent in a highly stylized way how financial and technical imperatives constrain a single fishing system that exploits a fish stock and sells its yield on a market. Such a fishing system could be a fishing country, its exclusive economic zone, its fleet and its fish market.
For ease of presentation, most of corresponding mathematical relationships are assumed to be linear. Relaxing this assumption does not affect our results qualitatively. Throughout, upper- and lower-case symbols denote variables and parameters respectively. 

The fishing system with a capacity $K$ for harvest  expends an effort $E$, which cannot exceed capacity ($E \leq K$), into harvesting a fish stock that has a total biomass $S$. The yield $Y$ from harvest depends on the level of stock $S$, as well as on effort $E$, according to the usual formula, 
$
Y = q S E, 
$
where $q$ is the fishing efficiency coefficient and represents the technological level of the fishing system. In turn, the biomass of fish stock $S$ decreases with yield $Y$,
$
S = r - s Y,
$
where $r$ denotes the level of biomass in the absence of fishing, and $s$ captures the effect of fishing on biomass. 

The profit of the fishing system depends on a balance of income and expenditures. Its only source of income is from sales. Its yield $Y$ is sold on the market, with unitary price $P$ decreasing with the total level of offer, 
$
P = a - b Y,
$ 
where $a$ is the maximum unitary price, and $b$ is the decrease in price according to offer. The system faces two sources of expenditures. First, fishing activity is costly. Unitary fishing costs decrease with stock according to:
$
 f = g + h / S
$. 
Secondly, access to the market is costly, with each unit of yield having a cost $m$. The net profit $I$ of the fishing system therefore is 
$ I = Y P - Y ( f + m)$. 

 From the above, we see that fish stock $S$, fish price $P$, income $I$ can be expressed in terms of yield $Y$ and  fishing capacity $K$. Henceforth, we will consider that yield and capacity are exogenous and underlie the behavior of the fishing system. The other variables are considered endogenous. 

In  conventional models of fisheries, the only -- and often left unmentioned -- constraint on a fishing system is that profit must be non-negative: $I \ge 0$. By contrast, in our framework, we will consider two constraints on the system. First, finance imposes an imperative rate of return upon the capital, $ K p $ where $p$ is the price of a fishing capacity unit of the fishing system. Financial constraint is expressed as: $ I \ge K p k$, where $k$ is the imperative rate of return, which implies an upper bound on capital: $ K \le I/(p k)$.  
Second, the capacity constraint $ E \le K $ can be re-written as $  Y / (q S)  \le K $.

\begin{figure}[h]
\begin{center}
\includegraphics[width=0.5\textwidth]{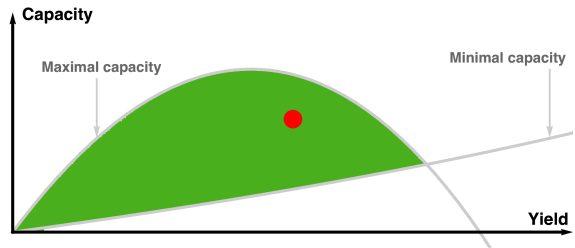}
\end{center}
\caption{ \textit{The space of feasible states for a fishing system. A fishing system is defined by its yield $Y$ (x-axis) and its  capacity $K$ (y-axis). The lower bound reflects the technical constraint: a minimal capacity is required to achieve a certain yield. The upper bound, meanwhile, is the financial constraint. Together, the constraints define a feasible state space where capacity allows yield and yield allows to finance capacity. Feasible states are shown in green. }}
\label{fig:fig1}
\end{figure}

\begin{figure}[h]
\begin{center}
\includegraphics[width=\textwidth]{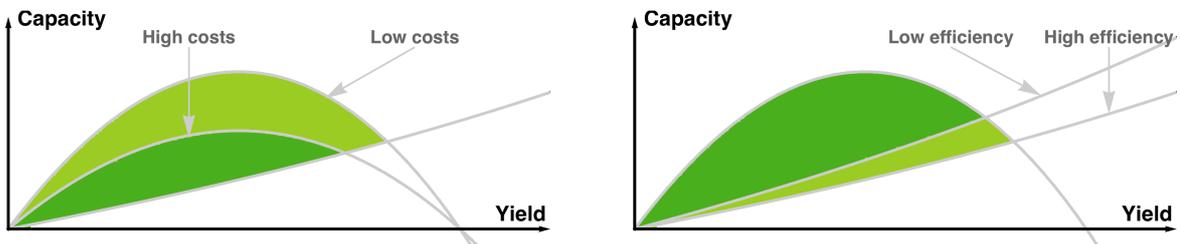}
\end{center}
\caption{\textit{Effects of fishing efficiency and rate of return on feasible space. Left: Decreasing fishing efficiency $q$  increases the minimal capacity and shrinks the feasible space. Right: Increasing rate of return $k$ decreases the maximum capacity and and shrinks the feasible space}}
\label{fig:fig2}
\end{figure}

If we combine both constraints, we obtain a space of feasible states in terms of the endogenous variables, capacity $K$ and yield $Y$,
\begin{equation} \label{eq:constraints}
E = \frac{Y}{q S}  \le K  \le \frac{Y(P -f-m)}{kp} = \frac{I}{pk},
\end{equation}
that is shown in figure \ref{fig:fig1}. With the idea that a system's behaviour is defined by capacity $K$ and yield $Y$, the size and shape of the space of feasible states constrain fishing behaviour. As shown in fig. \ref{fig:fig2}, the space of feasible states is particularly sensitive to finance constraints $k$ and fishing efficiency $q$ (i.e., technology).

\begin{figure}[h]
\begin{center}
\includegraphics[width=0.8\textwidth]{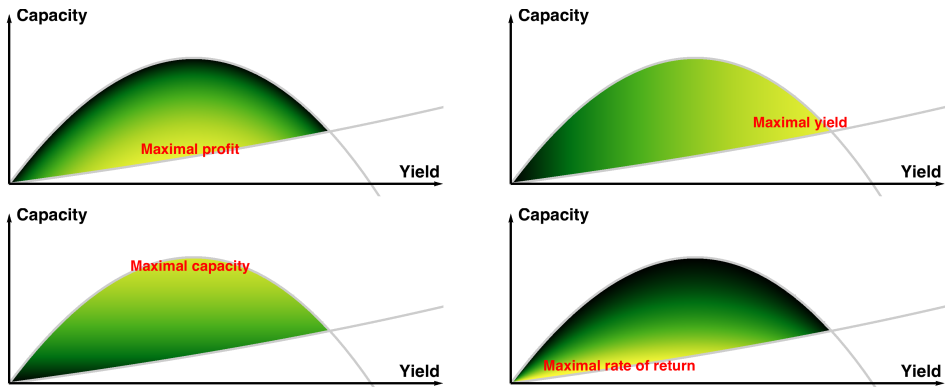}
\end{center}
\caption{ \textit{Constraints and rationality. The space of feasible state is colored (increasing from green to yellow) according to the values of variables (endogenous yield and capacity or exogenous profit and rate of return.} }
\label{fig:fig3}
\end{figure}

The position of a fishing system within its feasible state space, and the corresponding effect on other variables of this position, allows to infer on the underlying rationality of the system's behaviour. As shown in Fig~\ref{fig:fig3}, the position of a fishing system differs according to whether it seeks to maximize profit, capacity, yield or rate of return. In particular, if a fishing system is positioned close to the capacity boundary, then the corresponding fishing system is overcapacited and is not optimizing its profit because this requires being close to minimal capacity. If it is close to the left boundary of the feasible space, the system is overcapitalized, and again, it is not optimizing its profit. The explicit consideration of the constraints on fisheries therefore offers an intuitive framework to visualise important features of fishing system, such as excess capacity and overcapitalization.

\subsection{Two fishing systems in a constraint-based framework}

We now extend our constraint framework to two competing fishing systems, labelled $A$ and $B$, that exploit the same stock and sell their yield on the same market. $A$ and $B$ could be two countries with their own fleets that fish in the same offshore part of the ocean and sell their yield on the same international market. $A$ and $B$ respectively have capacities $K_A$ and $K_B$ and choose to expand an effort $E_A \leq K_A$ and $E_B \leq K_B$ into fishing. The yields of both systems depend on their respective efforts and fishing efficiencies ($q_A$ and $q_B$), $Y_A = q_A S E_A$ and $Y_B = q_B S E_B$, respectively. Natural fishing stock and unitary selling market price decrease with the yields of both fishing systems according to $ S =  r - s (Y_A+Y_B)$ and  $ P =  a - b (Y_A+Y_B)$. Fishing systems then differ in terms of fishing costs ($ f_A = g_A + h_A / S$, $ f_B = g_B + h_B / S$), marketing costs ($m_A$, $m_B$),  vessel values ($p_A$, $p_B$), and imperatives rates of return ($k_A$, $k_B$). The profits of $A$ and $B$ then are $ I_A = Y_A (P -  f_A - m_A)$ and $ I_B = Y_B (P- f_B - m_B)$, and must satisfy $I_A \geq K_A p_A k_A$ and $I_B \geq K_B p_B k_B$, respectively.

\begin{figure}[h]
\begin{center}
\includegraphics[width=\textwidth]{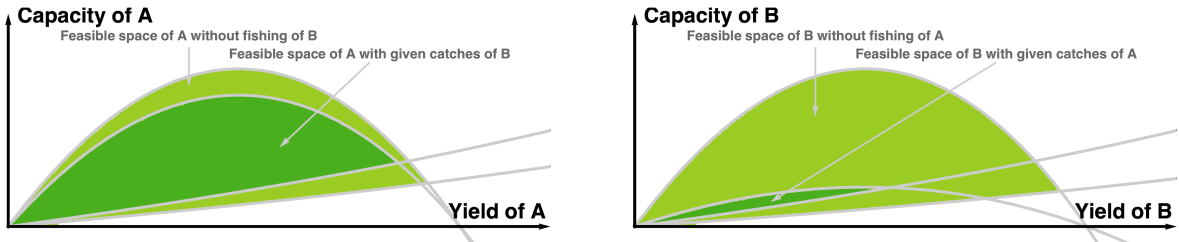}
\end{center}
\caption{\textit{Competition between fishing systems. The feasible set of a fishing system depends on the yield of the other fishing system} }
\label{fig:fig4}
\end{figure}

As in the preceding section, we combine the constraints of the framework to find the space of feasible states for each fishing system in terms of capacity and yield:
\begin{equation} \label{eq:constraints2}
\begin{split}
 \frac{Y_A}{q_A S} & \le K_A \le  \frac{Y_A(P-f_A-m_A)}{k_Ap_A}, \\
 \frac{Y_B}{q_B S} & \le K_B \le  \frac{Y_B(P-f_B-m_B)}{k_Bp_B}.
 \end{split}
\end{equation}
By contrast to the single fishing system case, fish stock $S$ and prices $P$ depend on total yield $ Y_A + Y_B$ here. Therefore the spaces of feasible states for each system are   inter-dependent through the yield of the other.  In particular, if one fishing system increases its yield, it shrinks the feasible space of the other (fig. \ref{fig:fig4}).

\subsection{The viability threshold}

 The space of feasible states entail that there is a maximum total yield ($Y_A$ for one fishing system, $Y_A+Y_B$ for two competing systems) that cannot be exceeded in order for fishing systems to survive. We define this value as the viability threshold and write it as $T$ (fig. \ref{fig:fig5}). Its derivation and expression, when there is a single fishing system, are given in appendix 1 (equation \ref{apdx:eq1}). 

If there is a single fishing system, the viability threshold depends negatively on the rate of return $k$ and positively on the fishing efficiency $q$, and on a balance between price and fixed costs: $a - m - g$ (figure \ref{fig:fig2}, see appendix 1, eqs. \ref{apdx:eq2}-\ref{apdx:eq4}). The existence of a viability threshold is due to finance and its relationship with technology: there is a value of yield that an system cannot exceed because otherwise, it is not able to meet the financial imperatives that are necessary to purchase the technology to generate that yield. 

If there are two fishing systems, their viability thresholds $T_A$ and $T_B$  depend on the properties of each system in the same way as when there is only one fishing system (eq. \ref{apdx:eq1} in appendix 1): the viability threshold of a system depends negatively on rates of return ($k_A$ or $k_B$) and positively on the fishing efficiency ($q_A$ or $q_B$) and on a balance of prices and costs ($a - m_A - g_A$ or $a - m_B - g_B$). This means  that the viability thresholds of two competing may be different. As a consequence, some amount of total yield may be viable for one system, but  simultaneously be the death warrant for the other (e.g., if $T_B < Y_A+Y_B < T_A$, then only $A$ can survive).

\subsection{Competition of two fishing systems within constraints}

Within these constraints, we now assume that fishing systems have a quantified objective for which they compete. This competition can be expressed as a non-cooperative game, in which players are fishing systems, strategies are pairs of yield and capacity, and payoffs are their quantified objective. As the set of feasible strategies of a player depends on the strategy of the other, such a game is said to be a Generalized Nash's equilibrium Problem \citep{facchinei2007generalized}. As usual, the equilibria of the game are pairs of strategies $\left( (Y_A,K_A), (Y_B,K_B) \right)$ such that when a player unilaterally changes its strategy, it obtains a smaller payoff \citep{facchinei2007generalized}.

We will study the game and its equilibria under two scenarios. In the first scenario, the quantified objective of the systems is profit. When $A$ and $B$ have equal viability thresholds ($T_A = T_B$), both systems stably co-exist (fig. \ref{fig:fig6}, top).  When $A$ and $B$ have different viability thresholds ($T_A \neq T_B$), the situation is more complicated (fig. \ref{fig:fig6}, middle and bottom). For the sake of argument, suppose the viability threshold of $A$ is greater than $B$'s ($T_A > T_B$), i.e., $A$ is in a dominant position and can exclude $B$ from the competition by choosing a yield greater than $B$'s viability threshold ($T_A > Y_A > T_B$). We find two possible outcomes, either $A$'s optimal (i.e., which maximises profit) yield is greater than $T_B$ in which case $B$ is excluded and dies out, or $A$'s optimal yield is lower than $T_B$ in which case $B$ survives (this is what is plotted in fig. \ref{fig:fig6}, right). The greater the difference between the viability thresholds of two competing systems, the more likely one system will vanish (fig. \ref{fig:fig6}, bottom). 

In the second scenario, the quantified objective of the systems is capacity. Under this scenario, we find that when $A$ and $B$ have equal viability thresholds, the game has an infinite number of equilibria (fig. \ref{fig:fig7}, top). Every sharing of the total yield, which equals the viability threshold of both systems ($Y_A + Y_B = T_A = T_B $), is a possible equilibrium. This process can be characterized by a Nash demand game \citep{nash1953two}. In order for them to stably co-exist, it is therefore for them to agree on how to share the total yield. Cooperation in this case is a requirement. When $A$ and $B$ have different viability thresholds, the game has a single equilibrium under which only the system with the greatest viability threshold survives (fig. \ref{fig:fig7}, bottom). This is because by maximizing its capacity, the system with the greatest viability threshold always pushes the total yield above the threshold of the other system, which can no longer satisfy its financial constraints and thus goes out of business.

\subsection{Coordination and bargaining of two fishing systems in a constraint-based framework}

In some cases, there exists pairs $( (Y_A,K_A), (Y_B,K_B))$ of strategies that are away from the Nash equilibrium and that provide a better reward to both players. Such strategies constitute the bargaining space of the system as both players have an interest to bargain in order to reach these strategies. Bargaining spaces are plotted in the right parts of figs \ref{fig:fig6} and \ref{fig:fig7}. Inside the bargaining there is what is referred to as the Pareto frontier. It is made of pairs of strategies such that an unilateral change of strategy results in a smaller reward to one one player (see \citep{muthoo1999bargaining} for definitions). The goal of bargaining is to reach a point on the Pareto frontier. 

As the top and centre right plots of fig.~\ref{fig:fig6} illustrate, when players maximize profit and have equal or marginally different viability thresholds, a bargaining space exists. Hence, competing players may reach an agreement through bargaining. This situation parallels the prisoner's dilemma problem: both players may reap greater payoffs if they coordinate to cooperate. When have important differences in viability thresholds, the bargaining space is empty (bottom right of fig.~\ref{fig:fig6}) is because the dominant player has no interest in bargaining: it can simply eliminate the other from the competition. Differences in viability thresholds therefore reduce the possibilities of coordination. 
 
When players maximize their capacity and have equal viability thresholds, the bargaining space is large (top right of fig.~\ref{fig:fig7}) because all pairs of possible strategies other than the Nash equilibrium provide a better reward. However, the situation is different to the case when players maximise profit because here, the threat values of a player (i.e., the maximum payoff it obtains in the absence of negotiation and with the other player’s worst strategy) is zero. Bargaining is therefore imperative and the situation is akin to a pure demand game in which players have to agree on the sharing of a pie \citep{nash1953two}. When players maximize capacity, any difference in viability thresholds leads to the collapse of the bargaining space (centre and bottom right of fig.~\ref{fig:fig7}). As before, this is because the dominant player can exclude the other from the competition and thus has no need to bargain. The sensitivity of the bargaining space to difference in viability thresholds when players maximize capacity highlights that when fishing systems operate
with modes of rationality other than profit maximization, possibilities for bargaining can be significantly destabilized.

Our analyses show that differences in viability thresholds compromise the co-existence of multiple systems, first by exacerbating the effects of competition and second by shrinking the possibilities of bargaining. This is particularly so when capacity is maximized. When systems have different viability thresholds and are maximizing capacity, two steps are necessary to ensure their long-term survival. First, fisheries management must aim at equalizing viability thresholds among competing systems. Recall that viability thresholds $T$ depend positively on cost related function $(a - m - f)$ and fishing efficiency $q$, and negatively on rates of returns $k$. There are therefore three ways to equalize viability thresholds. For the sake of argument, suppose that  $A$ is dominant over $B$, i.e., $T_B < T_A$. Then, to reach $T_B = T_A$, it is possible to modulate costs (i.e., increase $(a - m_B - f_B)$ and/or decrease $(a - m_A - f_A)$), to change fishing efficiencies (i.e., increase $B$'s $q_b$  and/or decrease $A$'s $q_A$), or change the rates of return (i.e.,  increase  $A$'s $k_A$ and/or  decrease $B$'s $k_B$). Second, players must find a fair division of the yield between them. 

\section{Discussion}

\subsection{Results summary}

We have framed a game between fishing systems within the technological and financial constraints they face. These constraints are reflected in the existence of \textit{viability thresholds}, which correspond to the maximum total fish resource that a fishing system can catch. A fishing system cannot exceed its viability threshold because otherwise, it is not able to meet the financial imperatives that are necessary to purchase the technology to generate that yield. Different fishing systems exploiting the same resources influence one another by constraining their scope of action, and when actors face different constraints, one can potentially ``kill" the other by pushing the system above the viability threshold of its competitor.

When we analysed games' equilibria and bargaining spaces, we found that they depend critically on viability thresholds and how they differ among players. When fishing systems have different viability thresholds, competition is exaggerated, and the transition towards coordination through bargaining that is needed to avoid over-exploitation and over-capacity is compromised. This is because when a dominant can simply eliminate the other from the competition, it has no motivation in finding a solution through bargaining.

\subsection{About fishing capacity maximization behavior}

We found that the effects of differences in viability thresholds are particularly important when fishing systems maximize capacity. This result leads to the question of the pertinence of such a behaviour, which may appear incongruous at first. The difference between profit maximisation and capacity maximisation as a goal can be framed as a question: at national level, do national fishing policies encourage investment in boats for insuring a bigger income to a country, or do they target a minimal profit to fishing companies for insuring the development of their national fishing fleet? Or, at another level, do fishing entities acquire big boats to catch fish and thus earn money, or do they sell fish in order to earn money and thus finance big boats? The answer to such questions is not obvious, at least to us.  

Capacity maximisation is perhaps the most parsimonious explanation to overinvestment and overcapacity \citep{bell2016global}. A more explicit illustration of fishing capacity maximisation can be found in deliberate national policies. In India for example, there have been 12 national plans since 1950 that all support capacity development  \citep{bhathal2014Government}, which has led to significant excess capacity, a production that exceeds the needs of the country \citep{bhathal2008fishing, indiaMarinFisheriesWB2010, sathyapalan2011overcapitalization, bhathal2014Government} and a national opposition to any international agreement for banning subsidies \citep{campling2013mainstreaming}. This nationwide increase of capacity was initially motivated by the necessity to develop and to eradicate poverty \citep{thorpe2005fisheries}, but seems today to be motivated by the anticipation of competition with fleets from developed countries \citep{campling2013mainstreaming}..

The notion that fishing entities favour an increase in capacity also provides a straightforward explanation to (1) the failures of decommissioning schemes (i.e., plans for decreasing fleets size through buybacks) even when they come at the expense of profit \citep{holland1999fishing}, (2) the observation that most subsidies, whatever their intentions, are redirected towards new equipment  \citep{sumaila2010bottom}, (3) the high technical turnover \citep[i.e., capital stuffing][]{townsend1985capital} that has been observed  in many fishing entities \citep{pauly2002towards}. Capacity maximisation in fisheries behaviour may be motivated by the anticipation of future capacity limitations \citep{guyader2001distributional}, of individual quotas distributed according to capacity \citep{squires1995individual}, or of subsidies of buybacks programs \citep{clark2005subsidies}, or by the desire to diminish the risk of exploiting a highly variable resource \citep{branch2006fleet}.

In addition, the standard hypothesis that fishing systems behave as to maximise profit does not always fit well with the empirical evidence.
Fisheries anthropology \citep{acheson1981anthropology, palsson1988models, olson1997cultural}, studies on fishermen behavior \citep{jentoft1989fisheries, salas2007small, holland2008fishermen, rijnsdorp2008arms} and on their communities \citep{jentoft2000community, pomeroy2006coping, bene2010not} describe fishermen who are mostly committed to their occupation, even if that means remaining poor, and who co-exist by relying on mutualistic rules \citep{jentoft1986fisheries, allison2001livelihoods}. They constitute closely-knit, sometimes closed, societies, within which competition is often symbolical: for the biggest fish, the biggest catch, the most productive fishing plots, the most efficient gear, the biggest boat or the best skipper \citep{palsson1988models}. 

Altogether, these observations lend support to the idea the some fishing systems are not only concerned with profit, but may also seek to increase capacity as a goal. The significant effects of capacity maximisation on constrained fisheries that we observed on our results call for future assessments of whether such behaviour is at play, and more in-depth analysis of its effects on competition and bargaining. 

\subsection{The future for ocean governance}

As economic development and population growth increase demand for seafood, technological progress is opening up new fishing areas in high- and deep-seas, which according to the 1982 United Nations Convention on the Law of the Sea (UNCLOS) are governed by a principle of freedom of fishing \citep{oda1983fisheries}. For reasons that are biological, ethical \citep{moore2016governing, barbier2014protect, brooks2013challenging} or economic \citep{rogers2014high}, a growing number of calls have been made to control access and organise negotiations on how to distribute fishing pressure in these new areas \citep{hayashi2004global, molenaar2005addressing, moore2016governing, barbier2014protect}. Future issues faced by such a global ocean governance and how they relate to our results can be understood by first considering the context in which they would take place. 

The institutional context is one in which governance depends on multilateral negotiations \citep{barrett2005theory}, which rely on a principle of efficiency (negotiation must succeed) and of rationality (players must have an interest in the success of the negotiation). Meanwhile, the economic context is one of globalization and financialization \citep{whalen2001integrating}, which in fishing systems generates highly competitive, highly efficient, excessively capitalized fishing entities \citep{squires2013technical, eigaard2014technological} due to a mutual reinforcement of technical progress, financial modernization and competitiveness. Finally, the international political context is underlain by a requirement for equity  \citep{ostrom1999revisiting,kaul1999defining, rao1999global, sen1999global}, which emphasizes that how the access to fish is shared should be fair, especially for developing countries \citep{allison2001big}.

In view of this context and our results, we may anticipate several possible futures for ocean governance. First, a politically stressed international context leads to the abandonment of the principle of equity and all obstacles to \textit{sea grabbing} are lifted.  The re-enforcement of technology, finance and competition continues and leads to a world in which only a small number of large fishing companies survive by providing revenues to the globalized finance system. Second, following what has been done for climate change negotiations, countries agree on an institutional frame but do not consider any kind of rationality other than profit maximization, and thus proceed to use tools such as side payments to ensure equity, which probably fail for the same reasons buybacks programs fail. Third, the re-enforcement of technology, finance and competition is halted by fisheries policies that (1) maintain fragile differentiated independent local finance systems, such as local cooperative credit systems, in order to emancipate fishing entities from globalized finance; and (2) stabilise fishing efficiency at a reasonable level, instead of systematically promoting competition by encouraging efficiency. As a result, viability thresholds are levelled, bargaining spaces are enlarged and the equity issue can be appropriately addressed.

\subsection{Conclusion}

To conclude, our model highlights that because the competition of fishing systems occurs within constraints, it is inherently unfair as differences in viability thresholds can lead to the survival of only the most viable fishing systems. This unfairness is exaggerated when fishing systems operate with modes of rationality other than profit maximization, such as capacity maximisation, that destabilize the possibilities for bargaining. Altogether, our results lead us to question the relevance of creating economical tools, such property rights, fishing rights or side payments, that aim towards enlarging the space of bargaining in the framework of the prisoner's dilemma, adding to the growing number of voices that question this framework \citep{Feeny1990Tragedy, alcock2002bargaining, richerson2002evolutionary}. In order to fulfill the demands of equity on future global governance, we suggest that it would be more relevant to focus on equalizing the constraints on fishing systems (i.e., their viability thresholds) by paying special attention to their local financial systems and by limiting technological development.


\section*{References}

\bibliographystyle{elsarticle-harv.bst}
\bibliography{allBiblio}

\begin{thebibliography}{75}
\expandafter\ifx\csname natexlab\endcsname\relax\def\natexlab#1{#1}\fi
\expandafter\ifx\csname url\endcsname\relax
  \def\url#1{\texttt{#1}}\fi
\expandafter\ifx\csname urlprefix\endcsname\relax\def\urlprefix{URL }\fi

\bibitem[{Acheson(1981)}]{acheson1981anthropology}
Acheson, J.~M., 1981. Anthropology of fishing. Annual review of anthropology,
  275--316.

\bibitem[{Alcock(2002)}]{alcock2002bargaining}
Alcock, F., 2002. Bargaining, uncertainty, and property rights in fisheries.
  World Politics 54~(04), 437--461.

\bibitem[{Allison(2001)}]{allison2001big}
Allison, E., 2001. Big laws, small catches: global ocean governance and the
  fisheries crisis. journal of International Development 13~(7), 933--950.

\bibitem[{Allison and Ellis(2001)}]{allison2001livelihoods}
Allison, E.~H., Ellis, F., 2001. The livelihoods approach and management of
  small-scale fisheries. Marine policy 25~(5), 377--388.

\bibitem[{Armstrong(1998)}]{armstrong1998sharing}
Armstrong, C.~W., 1998. Sharing a fish resource: Bargaining theoretical
  analysis of an applied allocation rule. Marine Policy 22~(2), 119--134.

\bibitem[{Arnason et~al.(2000)Arnason, Magnusson, and
  Agnarsson}]{arnason2000norwegian}
Arnason, R., Magnusson, G., Agnarsson, S., 2000. The norwegian spring-spawning
  herring fishery: a stylized game model. Marine Resource Economics, 293--319.

\bibitem[{Bailey et~al.(2010)Bailey, Sumaila, and
  Lindroos}]{bailey2010application}
Bailey, M., Sumaila, U.~R., Lindroos, M., 2010. Application of game theory to
  fisheries over three decades. Fisheries Research 102~(1), 1--8.

\bibitem[{Barbier et~al.(2014)Barbier, Moreno-Mateos, Rogers, Aronson,
  Pendleton, Danovaro, Henry, Morato, Ardron, Van~Dover,
  et~al.}]{barbier2014protect}
Barbier, E.~B., Moreno-Mateos, D., Rogers, A.~D., Aronson, J., Pendleton, L.,
  Danovaro, R., Henry, L.-A., Morato, T., Ardron, J., Van~Dover, C.~L., et~al.,
  2014. Protect the deep sea. Nature 505~(7484), 475--477.

\bibitem[{Barrett(2005)}]{barrett2005theory}
Barrett, S., 2005. The theory of international environmental agreements.
  Handbook of environmental economics 3, 1457--1516.

\bibitem[{Bell et~al.(2016)Bell, Watson, and Ye}]{bell2016global}
Bell, J.~D., Watson, R.~A., Ye, Y., 2016. Global fishing capacity and fishing
  effort from 1950 to 2012. Fish and Fisheries.

\bibitem[{B{\'e}n{\'e} et~al.(2010)B{\'e}n{\'e}, Hersoug, and
  Allison}]{bene2010not}
B{\'e}n{\'e}, C., Hersoug, B., Allison, E.~H., 2010. Not by rent alone:
  Analysing the pro-poor functions of small-scale fisheries in developing
  countries. Development Policy Review 28~(3), 325--358.

\bibitem[{Berkes et~al.(2008)Berkes, Colding, and Folke}]{berkes2008navigating}
Berkes, F., Colding, J., Folke, C., 2008. Navigating social-ecological systems:
  building resilience for complexity and change. Cambridge University Press.

\bibitem[{Bhathal(2014)}]{bhathal2014Government}
Bhathal, B., 2014. {Government led development of India's marine fisheries
  since 1950. Catch and effort trends, and bioeconomic models for exploring
  alternative policies}. Ph.D. thesis, The University of British Coumubia.

\bibitem[{Bhathal and Pauly(2008)}]{bhathal2008fishing}
Bhathal, B., Pauly, D., 2008. {Fishing down marine food webs and spatial
  expansion of coastal fisheries in India, 1950--2000}. Fisheries Research
  91~(1), 26--34.

\bibitem[{Binmore et~al.(1986)Binmore, Rubinstein, and
  Wolinsky}]{binmore1986nash}
Binmore, K., Rubinstein, A., Wolinsky, A., 1986. The {N}ash bargaining solution
  in economic modelling. The RAND Journal of Economics, 176--188.

\bibitem[{Branch et~al.(2006)Branch, Hilborn, Haynie, Fay, Flynn, Griffiths,
  Marshall, Randall, Scheuerell, Ward, et~al.}]{branch2006fleet}
Branch, T.~A., Hilborn, R., Haynie, A.~C., Fay, G., Flynn, L., Griffiths, J.,
  Marshall, K.~N., Randall, J.~K., Scheuerell, J.~M., Ward, E.~J., et~al.,
  2006. Fleet dynamics and fishermen behavior: lessons for fisheries managers.
  Canadian journal of Fisheries and Aquatic Sciences 63~(7), 1647--1668.

\bibitem[{Bromley(2009)}]{bromley2009abdicating}
Bromley, D.~W., 2009. Abdicating responsibility: the deceits of fisheries
  policy. Fisheries 34~(6), 280--290.

\bibitem[{Brooks et~al.(2013)Brooks, Weller, Gjerde, and
  Sumaila}]{brooks2013challenging}
Brooks, C.~M., Weller, J.~B., Gjerde, K., Sumaila, U.~R., 2013. Challenging the
  right to fish in a fast-changing ocean. Stan. Envtl. LJ 33, 289.

\bibitem[{Campling and Havice(2013)}]{campling2013mainstreaming}
Campling, L., Havice, E., 2013. {Mainstreaming environment and development at
  the World Trade Organization. Fisheries subsidies, the politics of
  rule-making and the elusive triple win}. Environment and Planning 45~(4),
  835--852.

\bibitem[{Clark et~al.(2005)Clark, Munro, and Sumaila}]{clark2005subsidies}
Clark, C.~W., Munro, G.~R., Sumaila, U.~R., 2005. Subsidies, buybacks, and
  sustainable fisheries. journal of Environmental Economics and Management
  50~(1), 47--58.

\bibitem[{Contreras et~al.(2004)Contreras, Klusch, and
  Krawczyk}]{contreras2004numerical}
Contreras, J., Klusch, M., Krawczyk, J.~B., 2004. Numerical solutions to
  {N}ash-{C}ournot equilibria in coupled constraint electricity markets. IEEE
  Transactions on Power Systems 19~(1), 195--206.

\bibitem[{Cournot(1838)}]{cournot1838recherches}
Cournot, A.-A., 1838. Recherches sur les principes math{\'e}matiques de la
  th{\'e}orie des richesses par Augustin Cournot. chez L. Hachette.

\bibitem[{Eigaard et~al.(2014)Eigaard, Marchal, Gislason, and
  Rijnsdorp}]{eigaard2014technological}
Eigaard, O.~R., Marchal, P., Gislason, H., Rijnsdorp, A.~D., 2014.
  Technological development and fisheries management. Reviews in Fisheries
  Science \& Aquaculture 22~(2), 156--174.

\bibitem[{Epstein(2005)}]{epstein2005financialization}
Epstein, G.~A., 2005. Financialization and the world economy. Edward Elgar
  Publishing.

\bibitem[{Facchinei et~al.(2007)Facchinei, Fischer, and
  Piccialli}]{facchinei2007generalized}
Facchinei, F., Fischer, A., Piccialli, V., 2007. On generalized {N}ash games
  and variational inequalities. Operations Research Letters 35~(2), 159--164.

\bibitem[{Fearon(1998)}]{fearon1998bargaining}
Fearon, J.~D., 1998. Bargaining, enforcement, and international cooperation.
  International organization 52~(02), 269--305.

\bibitem[{Feeny et~al.(1990)Feeny, Berkes, Mccay, and
  Acheson}]{Feeny1990Tragedy}
Feeny, D., Berkes, F., Mccay, B., Acheson, J., 1990. The tragedy of the
  commons: Twenty-two years later. Human Ecology 18~(1), 1--19.

\bibitem[{Feeny et~al.(1996)Feeny, Hanna, and McEvoy}]{feeny1996questioning}
Feeny, D., Hanna, S., McEvoy, A.~F., 1996. Questioning the assumptions of the"
  tragedy of the commons" model of fisheries. Land Economics, 187--205.

\bibitem[{Gordon(1954)}]{gordon1954economic}
Gordon, H., 1954. The economic theory of a common-property resource: the
  fishery. The journal of Political Economy 62~(2), 124--142.

\bibitem[{Guyader and Th{\'e}baud(2001)}]{guyader2001distributional}
Guyader, O., Th{\'e}baud, O., 2001. Distributional issues in the operation of
  rights-based fisheries management systems. Marine Policy 25~(2), 103--112.

\bibitem[{Hardin(1968)}]{Hardin1968Tragedy}
Hardin, G., 1968. The tragedy of the commons. Science 162~(3859), 1243--1248.

\bibitem[{Hayashi(2004)}]{hayashi2004global}
Hayashi, M., 2004. Global governance of deep-sea fisheries. The International
  Journal of Marine and Coastal Law 19~(3), 289--298.

\bibitem[{Holland et~al.(1999)Holland, Gudmundsson, and
  Gates}]{holland1999fishing}
Holland, D., Gudmundsson, E., Gates, J., 1999. Do fishing vessel buyback
  programs work: a survey of the evidence. Marine Policy 23~(1), 47--69.

\bibitem[{Holland(2008)}]{holland2008fishermen}
Holland, D.~S., 2008. Are fishermen rational? a fishing expedition. Marine
  Resource Economics 23, 325--344.

\bibitem[{Jentoft(1986)}]{jentoft1986fisheries}
Jentoft, S., 1986. Fisheries co-operatives: lessons drawn from international
  experiences. Canadian journal of Development Studies/Revue canadienne
  d'{\'e}tudes du d{\'e}veloppement 7~(2), 197--209.

\bibitem[{Jentoft(1989)}]{jentoft1989fisheries}
Jentoft, S., 1989. Fisheries co-management: delegating government
  responsibility to fishermen's organizations. Marine policy 13~(2), 137--154.

\bibitem[{Jentoft(2000)}]{jentoft2000community}
Jentoft, S., 2000. The community: a missing link of fisheries management.
  Marine policy 24~(1), 53--60.

\bibitem[{Kaitala and Lindroos(2007)}]{kaitala2007game}
Kaitala, V., Lindroos, M., 2007. Game theoretic applications to fisheries. In:
  Handbook of Operations Research in Natural Resources. Springer, pp. 201--215.

\bibitem[{Kaul et~al.(1999)Kaul, Grunberg, and Stern}]{kaul1999defining}
Kaul, I., Grunberg, I., Stern, M.~A., 1999. Defining global public goods. In:
  Kaul, I., Grunberg, I., Stern, M.~A. (Eds.), Global Public Goods. Oxford
  University Press, pp. 68--87.

\bibitem[{Krawczyk(2005)}]{krawczyk2005coupled}
Krawczyk, J.~B., 2005. Coupled constraint {N}ash equilibria in environmental
  games. Resource and Energy Economics 27~(2), 157--181.

\bibitem[{Leonard(1994)}]{leonard1994reading}
Leonard, R.~J., 1994. Reading {C}ournot, reading {N}ash: The creation and
  stabilisation of the {N}ash equilibrium. The Economic Journal, 492--511.

\bibitem[{Li and Marden(2010)}]{li2010designing}
Li, N., Marden, J.~R., 2010. Designing games to handle coupled constraints. In:
  49th IEEE Conference on Decision and Control (CDC). IEEE, pp. 250--255.

\bibitem[{Lindroos and Kaitala(2000)}]{lindroos2000nash}
Lindroos, M., Kaitala, V., 2000. Nash equilibria in a coalition game of the
  norwegian spring-spawning herring fishery. Marine Resource Economics,
  321--339.

\bibitem[{Lindroos et~al.(2005)Lindroos, Kronbak, and
  Kaitala}]{lindroos2005coalition}
Lindroos, M., Kronbak, L.~G., Kaitala, V., 2005. Coalition games in fisheries
  economics. In: Advances in Fisheries Economics: Festschrift in Honour of
  Professor Gordon R. Munro. Blackwell Publishing Ltd, pp. 184--195.

\bibitem[{Molenaar(2005)}]{molenaar2005addressing}
Molenaar, E.~J., 2005. Addressing regulatory gaps in high seas fisheries. The
  International Journal of Marine and Coastal Law 20~(3), 533--570.

\bibitem[{Moore and Squires(2016)}]{moore2016governing}
Moore, S., Squires, D., 2016. Governing the depths: Conceptualizing the
  politics of deep sea resources. Global Environmental Politics.

\bibitem[{Munro(2009)}]{munro2009game}
Munro, G.~R., 2009. Game theory and the development of resource management
  policy: the case of international fisheries. Environment and Development
  Economics 14~(01), 7--27.

\bibitem[{Muthoo(1999)}]{muthoo1999bargaining}
Muthoo, A., 1999. Bargaining theory with applications. Cambridge University
  Press.

\bibitem[{Myerson(1999)}]{myerson1999nash}
Myerson, R.~B., 1999. {N}ash equilibrium and the history of economic theory.
  journal of Economic Literature 37, 1067--1082.

\bibitem[{Nash(1950{\natexlab{a}})}]{nash1950bargaining}
Nash, J., 1950{\natexlab{a}}. The bargaining problem. Econometrica: journal of
  the Econometric Society, 155--162.

\bibitem[{Nash(1950{\natexlab{b}})}]{nash1950equilibrium}
Nash, J., 1950{\natexlab{b}}. Equilibrium points in n-person games. Proc. Nat.
  Acad. Sci. USA 36~(1), 48--49.

\bibitem[{Nash(1953)}]{nash1953two}
Nash, J., 1953. Two-person cooperative games. Econometrica: journal of the
  Econometric Society, 128--140.

\bibitem[{Norse et~al.(2012)Norse, Brooke, Cheung, Clark, Ekeland, Froese,
  Gjerde, Haedrich, Heppell, Morato, et~al.}]{norse2012sustainability}
Norse, E.~A., Brooke, S., Cheung, W.~W., Clark, M.~R., Ekeland, I., Froese, R.,
  Gjerde, K.~M., Haedrich, R.~L., Heppell, S.~S., Morato, T., et~al., 2012.
  Sustainability of deep-sea fisheries. Marine policy 36~(2), 307--320.

\bibitem[{Oda(1983)}]{oda1983fisheries}
Oda, S., 1983. Fisheries under the united nations convention on the law of the
  sea. The American Journal of International Law 77~(4), 739--755.

\bibitem[{Olson(1997)}]{olson1997cultural}
Olson, J.~A., 1997. The cultural politics of fishing: negotiating community and
  common property in Northern Norway. Stanford University.

\bibitem[{Ostrom et~al.(1999)Ostrom, Burger, Field, Norgaard, and
  Policansky}]{ostrom1999revisiting}
Ostrom, E., Burger, J., Field, C., Norgaard, R., Policansky, D., 1999.
  Revisiting the commons: local lessons, global challenges. science 284~(5412),
  278--282.

\bibitem[{Palsson(1988)}]{palsson1988models}
Palsson, G., 1988. Models for fishing and models of success. Maritime
  Anthropological Studies 1~(1), 15--28.

\bibitem[{Pauly et~al.(2002)Pauly, Christensen, Gu{\'e}nette, Pitcher, Sumaila,
  Walters, Watson, and Zeller}]{pauly2002towards}
Pauly, D., Christensen, V., Gu{\'e}nette, S., Pitcher, T., Sumaila, U.,
  Walters, C., Watson, R., Zeller, D., 2002. Towards sustainability in world
  fisheries. Nature 418~(6898), 689--695.

\bibitem[{Pintassilgo(2003)}]{pintassilgo2003coalition}
Pintassilgo, P., 2003. A coalition approach to the management of high seas
  fisheries in the presence of externalities. Natural Resource Modeling 16~(2),
  175--197.

\bibitem[{Pomeroy et~al.(2006)Pomeroy, Ratner, Hall, Pimoljinda, and
  Vivekanandan}]{pomeroy2006coping}
Pomeroy, R.~S., Ratner, B.~D., Hall, S.~J., Pimoljinda, J., Vivekanandan, V.,
  2006. Coping with disaster: Rehabilitating coastal livelihoods and
  communities. Marine Policy 30~(6), 786--793.

\bibitem[{Rao(1999)}]{rao1999global}
Rao, J.~M., 1999. Equity in a global publics good framework. In: Kaul, I.,
  Grunberg, I., Stern, M.~A. (Eds.), Global Public Goods. Oxford University
  Press, pp. 2--19.

\bibitem[{Richerson et~al.(2002)Richerson, Boyd, and
  Paciotti}]{richerson2002evolutionary}
Richerson, P.~J., Boyd, R., Paciotti, B., 2002. An evolutionary theory of
  commons management. The drama of the commons, 403--442.

\bibitem[{Rijnsdorp et~al.(2008)Rijnsdorp, Poos, Quirijns, HilleRisLambers,
  De~Wilde, and Den~Heijer}]{rijnsdorp2008arms}
Rijnsdorp, A.~D., Poos, J.~J., Quirijns, F.~J., HilleRisLambers, R., De~Wilde,
  J.~W., Den~Heijer, W.~M., 2008. The arms race between fishers. journal of Sea
  Research 60~(1), 126--138.

\bibitem[{Rogers et~al.(2014)Rogers, Sumaila, Hussain, and
  Baulcomb}]{rogers2014high}
Rogers, A., Sumaila, U., Hussain, S., Baulcomb, C., 2014. The high seas and us:
  Understanding the value of high-seas ecosystems. Global Ocean Commission.

\bibitem[{Salas and Charles(2007)}]{salas2007small}
Salas, S., Charles, A., 2007. Are small-scale fishers profit maximizers?:
  Exploring fishing performance of small-scale fishers and factors determining
  catch rates. Proceedings of the 60th Gulf and Caribbean Fisheries Institute
  60, 116--124.

\bibitem[{Sathyapalan et~al.(2011)Sathyapalan, Srinivasan, and
  Scholtens}]{sathyapalan2011overcapitalization}
Sathyapalan, J., Srinivasan, J.~T., Scholtens, J., 2011. Overcapitalization in
  a small-scale trawler fishery: a study of {P}alk {B}ay, {I}ndia. In: World
  Small-scale Fisheries: Contemporary Visions. Eburon Uitgeverij BV, pp.
  51--62.

\bibitem[{Sen(1999)}]{sen1999global}
Sen, A., 1999. Global justice: beyond international justice. In: Kaul, I.,
  Grunberg, I., Stern, M.~A. (Eds.), Global Public Goods. Oxford University
  Press, pp. 116--125.

\bibitem[{Squires et~al.(1995)Squires, Kirkley, and
  Tisdell}]{squires1995individual}
Squires, D., Kirkley, J., Tisdell, C.~A., 1995. Individual transferable quotas
  as a fisheries management tool. Reviews in Fisheries Science 3~(2), 141--169.

\bibitem[{Squires and Vestergaard(2013)}]{squires2013technical}
Squires, D., Vestergaard, N., 2013. Technical change in fisheries. Marine
  Policy 42, 286--292.

\bibitem[{Sumaila(1999)}]{sumaila1999review}
Sumaila, U.~R., 1999. A review of game-theoretic models of fishing. Marine
  policy 23~(1), 1--10.

\bibitem[{Sumaila et~al.(2010)Sumaila, Khan, Dyck, Watson, Munro, Tydemers, and
  Pauly}]{sumaila2010bottom}
Sumaila, U.~R., Khan, A.~S., Dyck, A.~J., Watson, R., Munro, G., Tydemers, P.,
  Pauly, D., 2010. A bottom-up re-estimation of global fisheries subsidies.
  journal of Bioeconomics 12~(3), 201--225.

\bibitem[{Thorpe et~al.(2005)Thorpe, Reid, van Anrooy, and
  Brugere}]{thorpe2005fisheries}
Thorpe, A., Reid, C., van Anrooy, R., Brugere, C., 2005. When fisheries
  influence national policy-making: an analysis of the national development
  strategies of major fish-producing nations in the developing world. Marine
  Policy 29~(3), 211--222.

\bibitem[{Townsend(1985)}]{townsend1985capital}
Townsend, R.~E., 1985. On "capital-stuffing" in regulated fisheries. Land
  Economics 61~(2), 195--197.

\bibitem[{Whalen(2001)}]{whalen2001integrating}
Whalen, C.~J., 2001. Integrating {S}chumpeter and {K}eynes: {H}yman {M}insky's
  theory of capitalist development. journal of Economic issues 35~(4),
  805--823.

\bibitem[{World-Bank(2010)}]{indiaMarinFisheriesWB2010}
World-Bank, 2010. India Marine Fisheries Issues, Opportunities and Transitions
  for Sustainable Development. Vol.~91. World Bank. Agriculture and Rural
  Development Sector Unit South Asia region.

\end{thebibliography}

\section{Appendix 1: Computing viability threshold \label{proofs}}
We show here that there exists a maximal value for yield and give its expression. We start from the technological and finance constraints on yield that have been obtained in above equation \ref{eq:constraints}:
\begin{eqnarray*}
E = \frac{Y}{q S(Y)}   & \le  K  & \le  \frac{Y(P(Y)-f(Y)-m)}{kp}  
\end{eqnarray*}
Using definitions: $S(Y) = r-sY$, $P(Y)=a-bY$ and $f(Y)=g+h/S(Y)$, they can be developed as:
\begin{eqnarray*}  
\frac{1}{q (r-sY)}   & \le   &   \frac{1}{kp} \times ((a-bY)-(g + \frac{h }{ r - s Y})-m)
\end{eqnarray*}
That is:
\begin{eqnarray*}  
kp & \le   &  ((a-bY)-(g + \frac{h }{ r - s Y})-m) (q (r-sY))
\end{eqnarray*}

To simplify, we put $ l  =  a - g - m$. Obviously situations with $l < 0$ are meaningless.  It is routine to show that previous inequality can be re-written as:
\begin{eqnarray} \label{apdx:eq0}
 A Y^2 + B Y + C & \ge & 0 
\end{eqnarray}
With  
\begin{eqnarray*}
A & = & bqs \\
B & = &  (-b q r - a q s + g q s + m q s) =   - q (br + sl) \le 0 \\
C & = & (-k p - h q + q r l)  = q ( r l - h ) -  k p
 \end{eqnarray*}

We look for values of $Y$ satisfying inequality \ref{apdx:eq0}. We solve the corresponding second degree equation. Possible yields if they exist are outside its roots. Its discriminant is:
\begin{eqnarray*}
\Delta & = & q^2 ( b r + s  l)^2 - 4 q b s (q  l r  - (q  h + k p )) \\ & = & q^2 ( b r - s  l)^2 + 4 q b s   (q  h + k p ) > 0
\end{eqnarray*}
Solutions are:
\begin{eqnarray*}
T_1 =  \frac{q (br + sl)  - \sqrt{\Delta}}{2 bqs } \\
T_2 =  \frac{q (br + sl)  + \sqrt{\Delta}}{2 bqs } 
 \end{eqnarray*}

Recall that $ S = r - s Y$. Thus we are interested only in values of $Y$ that are outside roots and such that $ Y \le r/s $. We remark that we have:
\begin{eqnarray*}
\sqrt{\Delta} = \sqrt{(br-sl)^2 + 4 q b s   (q  h + k p )} >  | br -sl |
\end{eqnarray*}
We deduce:
\begin{eqnarray*} 
 \frac{r}{s} - T_1 & = & \frac{br-sl}{2 bs}  + \frac{\sqrt{\Delta}}{2 bs}   >  0  \\
T_2 - \frac{r}{s} & = & \frac{- br+sl}{2 bs}  + \frac{\sqrt{\Delta}}{2 bs}  >  0 
\end{eqnarray*} 

Therefore, values of $Y$ greater than $T_2$ are not valid. Possible values for $Y$ are thus such that $ Y \le T_1$. Values of parameters $(a, b, s, r, g, h, etc.)$ resulting in a system with $ T_1 < 0 $ are not realistic.

Finally the viability threshold is:
\begin{eqnarray}
T & =  & \frac{q (br + sl)  - \sqrt{\Delta}}{2 bqs } \label{apdx:eq1}
\end{eqnarray}

It is easy to derive the dependency of the viability threshold according to systems parameters. We have:
\begin{eqnarray}
\frac{\partial T}{\partial q} & = & \frac{k p}{q \sqrt{\Delta}  } >  0 \label{apdx:eq2} \\
\frac{\partial T}{\partial k} & = & - \frac{p}{\sqrt{\Delta}  } < 0  \label{apdx:eq3} \\
\frac{\partial T}{\partial l} & = &  \frac{1}{2b} \frac{\sqrt{\Delta}  +  q(b  r - l  s)}{\sqrt{\Delta }  } 
 \ge  \frac{1}{2b} \frac{\sqrt{\Delta }  - | q(b  r - l  s) | }{\sqrt{\Delta }  }  > 0 \label{apdx:eq4}
\end{eqnarray}

\clearpage

\section*{Tables}

\begin{table}[h]
\centering\footnotesize\setstretch{1.5}
        \begin{tabular}{p{0.22\textwidth}p{0.34\textwidth}p{0.34\textwidth}}
\hline
& Maximizing profit & Maximizing capacity \\ \hline
Equal viability thresholds & One equilibrium with co-existence of both fishing systems. \newline An explorable bargaining space &  Every share of the total yield, which equals the viability threshold, is a Nash equilibrium. \newline The bargaining space is made of all possible pairs of strategies. The Pareto frontier is made of a sharing of the yield at the common value of viability thresholds. \\ \hline 
Sligthly different viability thresholds & One equilibrium which allows the co-existence of both systems. \newline An explorable, but complicated, bargaining space. 
& One equilibrium with the survival of only the dominant system. \newline Bargaining space is empty. \\ \hline
Different viability thresholds & One equilibrium with the survival of only the dominant system.  \newline Bargaining space is empty. 
& One equilibrium with the survival of only the dominant system. \newline Bargaining space is empty. \\ \hline
\end{tabular}
\caption{\normalsize{Competition between fishing systems. Nash equilibria and bargaining space.}}
\label{atvblef}
\end{table}

\clearpage

\section*{Figures captions}

\begin{figure}[h]
\begin{center}
\includegraphics[width=\textwidth]{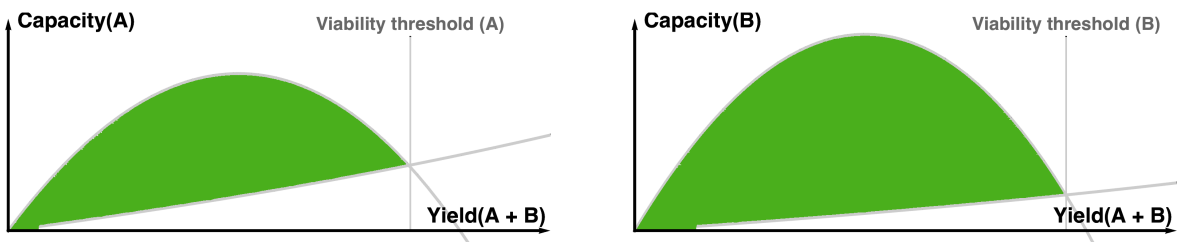}
\end{center}
\caption{ \textit{Viability thresholds of two competing fishing systems} }
\label{fig:fig5}
\end{figure}

\begin{figure}[h]
\includegraphics[width=\textwidth]{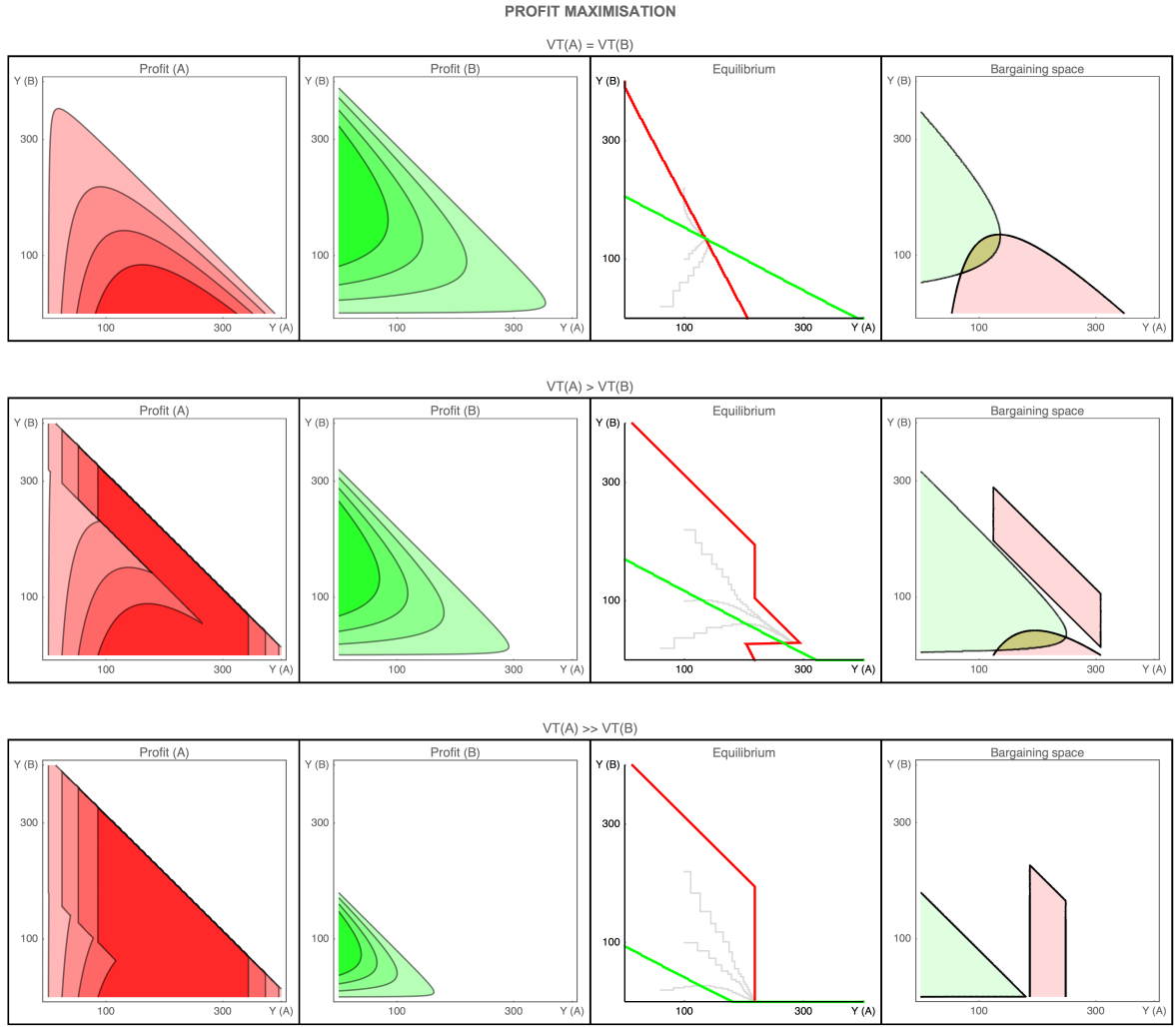}
\caption{\textit{Competitive equilibrium and bargaining space with profit maximisation.  From top to bottom row: (a) equal viability thresholds: $ VT(A) = VT(B) = 400$, (b) slightly different viability thresholds: $ VT(A) = 400, VT(B) = 300$, (c) different viability thresholds $ VT(A) = 400, VT(B) = 200$. From left to right column:  (a)  how profit of $A$ depends on yields $Y_A$ and $ Y_B$; white areas corresponds to infeasible states (total yields exceeds the viability threshold of $A$);  (b) how profit of $B$ depends on yields $Y_A$ and $ Y_B$; (c) Nash equilibrium of the competitive game (red line, the best strategy of $A$ according to $B$'s strategy; green line, the best strategy of $B$ according to $A$'s strategy; gray lines show how equilibrium can be reached  by successive adjustments corresponding to different starting points); (d) the bargaining space lies at the intersection of the green area (where $A$ get a better reward than at Nash's equilibrium) and the red area (where $A$ get a better reward than at Nash's equilibrium).}}
\label{fig:fig6}
\end{figure}

\begin{figure}[!h]
\begin{center}
\includegraphics[width=\textwidth]{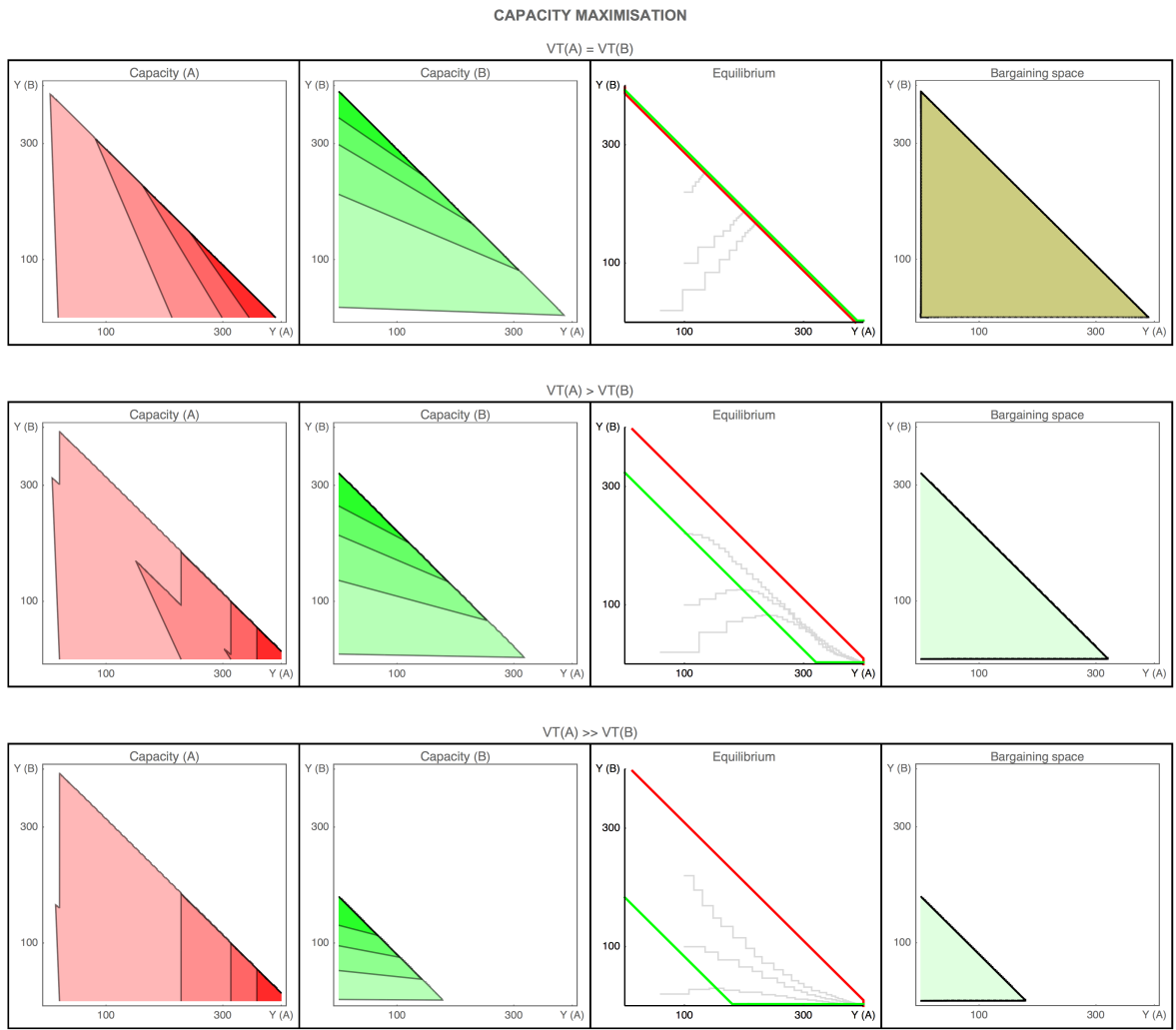}
\end{center}
\caption{\textit{Competitive equilibrium and bargaining space with capacity maximisation.  Similar to figure \ref{fig:fig6}. With equal viability thresholds, the bargaining space is mad of all possibles pairs of strategies. When they are different, it is empty. }}
\label{fig:fig7}
\end{figure}

\end{document}